\documentclass[prl,aps,floats,amsmath,amssymb,superscriptaddress,preprintnumbers,nofootinbib]{revtex4}
\usepackage{graphicx,floatflt,epsfig,epsf}

\def\be{\begin{equation}}
\def\ee{\end{equation}}
\def\bea{\begin{eqnarray}}
\def\eea{\end{eqnarray}}

\def\bbuildrel#1_#2^#3{\mathrel{\mathop{\kern 0pt#1}\limits_{#2}^{#3}}}
\def\slash#1{\setbox0=\hbox{$#1$}#1\hskip-\wd0\dimen0=5pt\advance
       \dimen0 by-\ht0\advance\dimen0 by\dp0\lower0.5\dimen0\hbox
         to\wd0{\hss\sl/\/\hss}}

\newcommand{\gae}{\lower 2pt \hbox{$\, \buildrel {\scriptstyle >}\over {\scriptstyle
\sim}\,$}}
\newcommand{\lae}{\lower 2pt \hbox{$\, \buildrel {\scriptstyle <}\over {\scriptstyle
\sim}\,$}}

\newcommand{\lt}{\left}
\newcommand{\rt}{\right}

\def\issue(#1,#2,#3){{\bf #1}, #2 (#3)}
\def\opcit(#1){ {\em op. cit.}, #1}

\def\APP(#1,#2,#3){Acta Phys.\ Polon.\ \issue(#1,#2,#3)}
\def\ARNPS(#1,#2,#3){Ann.\ Rev.\ Nucl.\ Part.\ Sci.\ \issue(#1,#2,#3)}
\def\CPC(#1,#2,#3){Comp.\ Phys.\ Comm.\ \issue(#1,#2,#3)}
\def\CIP(#1,#2,#3){Comput.\ Phys.\ \issue(#1,#2,#3)}
\def\EPJC(#1,#2,#3){Eur.\ Phys.\ J.\ C\ \issue(#1,#2,#3)}
\def\EPJD(#1,#2,#3){Eur.\ Phys.\ J. Direct\ C\ \issue(#1,#2,#3)}
\def\IEEETNS(#1,#2,#3){IEEE Trans.\ Nucl.\ Sci.\ \issue(#1,#2,#3)}
\def\IJMP(#1,#2,#3){Int.\ J.\ Mod.\ Phys. \issue(#1,#2,#3)}
\def\JHEP(#1,#2,#3){J.\ High Energy Physics \issue(#1,#2,#3)}
\def\JPG(#1,#2,#3){J.\ Phys.\ G \issue(#1,#2,#3)}
\def\MPL(#1,#2,#3){Mod.\ Phys.\ Lett.\ \issue(#1,#2,#3)}
\def\NP(#1,#2,#3){Nucl.\ Phys.\ \issue(#1,#2,#3)}
\def\NIM(#1,#2,#3){Nucl.\ Instrum.\ Meth.\ \issue(#1,#2,#3)}
\def\PL(#1,#2,#3){Phys.\ Lett.\ \issue(#1,#2,#3)}
\def\PRD(#1,#2,#3){Phys.\ Rev.\ D \issue(#1,#2,#3)}
\def\PRL(#1,#2,#3){Phys.\ Rev.\ Lett.\ \issue(#1,#2,#3)}
\def\SJNP(#1,#2,#3){Sov.\ J. Nucl.\ Phys.\ \issue(#1,#2,#3)}
\def\ZPC(#1,#2,#3){Zeit.\ Phys.\ C \issue(#1,#2,#3)}

\begin{document}

%\preprint{
%\vbox{
%\hbox{BNL-HET-08/17, TIFR/TH/08-29, HRI-P-08-07-001}
%}}

\title{The fourth family: a simple explanation for the observed pattern of 
anomalies in $B$-{\it CP} asymmetries}

\author{Amarjit Soni}
\affiliation{Physics Department, Brookhaven National Laboratory,
  Upton, NY 11973, USA}
\author{Ashutosh Kumar Alok}
\affiliation{Tata Institute of Fundamental Research, Homi Bhabha Road, Mumbai
400005, India}
\author{Anjan Giri}
\affiliation{Department of Physics, Punjabi University, Patiala-147002, 
  India}
\author{Rukmani Mohanta}
\affiliation{School of Physics, University of
  Hyderabad, Hyderabad - 500046, India}
\author{Soumitra Nandi}
\affiliation{Harish Chandra Research Institute,
Chhatnag Road, Jhusi, Allahabad- 211 019, India and Dipartimento di Fisica 
Teorica, Univ. di Torino and INFN, Sezione di Torino, I-10125 Torino, Italy}

\begin{abstract}
We show that a fourth family of quarks with $m_{t'}$ in the range of
( 400 - 600) 
GeV provides a rather simple explanation for the several indications of new 
physics that have been observed involving CP asymmetries of the b-quark. The 
built-in hierarchy of the 4$\times$4 mixing matrix is such that the $t'$ 
readily provides a needed {\it perturbation} ($\approx 15\%$) to $\sin 2 \beta$ as measured in
$B \to \psi K_s$ and simultaneously is the
dominant source of CP asymmetry in $B_s \to \psi \phi$.
%The difficulty
%in understanding the large observed difference
%in direct CP
%asymmetries in $\bar B^0 \to K^- \pi^+$ versus $B^- \to K^- \pi^0$ 
%also tends to get amiliorated.  
The correlation between CP asymmetries in $B_s \to
\psi \phi$ and $B_d\to\phi K_s$ suggests
$m_{t'} \approx$ (400 - 600) GeV. Such heavy masses point to the tantalizing 
possibility that the 4th family plays an important role in the electroweak 
symmetry breaking.
%as the Pagels-Stokar relation in fact requires quarks of 
%masses around 500 or 600 GeV for dynamical mass generation to take place.

\end{abstract}

\maketitle

The spectacular performance of the two asymmetric $B$-factories allowed us to 
reach an important milestone in our understanding of {\it CP}-violation phenomena. For the first time it was established that the observed {\it CP}-violation in
the $B$ and $K$ systems was indeed accountable by the single, {\it CP}-odd, 
Kobayashi-Maskawa phase in the CKM matrix \cite{1963, 1973}. In particular, the 
time dependent {\it CP}-asymmetry in the gold-plated $B^0\to\psi K_s$ can be 
accounted for by the Standard Model (SM) CKM-paradigm to an accuracy of around 
15\% \cite{HFAG, RMP}. It has then become clear that the effects of a beyond 
the standard model (BSM) phase can only be a perturbation. Nevertheless, in the 
past few years as more data were accumulated and also as the accuracy in some 
theoretical calculations was improved it has become increasingly apparent that 
several of the experimental results are difficult to reconcile within the SM 
with three generations [SM3] \cite{LS07,LS08}. It is clearly
important to follow these indications and to try to identify the possible origin
of these discrepancies especially since they may provide experimental signals
for the LHC which is set to start quite soon.
While at this stage many extensions of the SM could be responsible, in this 
paper, we will make the case that an addition of a fourth family of quarks~\cite{AS_olds1,AS_olds2,AS_olds3,AS_olds4,hung-sher} 
provides a rather simple explanation for the pattern of deviations that have 
been observed \cite{CHOU}. In fact we will show that the data suggests that the 
charge 2/3 quark of this family needs to have a mass in the range of (400 - 600) GeV\cite{error_etc}

We now briefly mention the experimental observations involving $B$-{\it CP}
asymmetries that are indicative of possible difficulties for the CKM picture of
{\it CP}-violation.
\begin{enumerate}
\item{The predicted value of $\sin 2\beta$ in the SM seems to be about 2-3 
$\sigma$ larger than the directly measured values. Using only $\epsilon_k$ and 
$\Delta M_s/\Delta M_d$ from experiment along with the necessary hadronic 
matrix elements, namely kaon ``$B$-parameter" $B_K$ and using $SU(3)$ breaking
ratio $\xi_s \equiv {f_{bs}\sqrt{B_{bs}}\over f_{bd}\sqrt{B_{bd}}}$, from the
lattice, alongwith $V_{cb}$ yields a prediction, $\sin 2\beta^{prediction}_{no
V_{ub}} = 0.87 \pm 0.09$ \cite{LS08} in the SM. If along with that $V_{ub}\over V_{cb}$ is 
also included as an input then one gets a somewhat smaller central value but 
with also appreciably reduced error: $\sin 2\beta^{prediction}_{full fit} = 
0.75 \pm 0.04$. }
\item{The celebrated measurement, via the ``gold-plated" mode $B\to \psi K_s$, 
gives $\sin 2\beta_{\psi K_s} = 0.672 \pm 0.024$ which is smaller than either
of the above predictions by $\approx$ 1.7 to 2.1 $\sigma$ \cite{LS08}.}

\item{As is well known penguin-dominated modes, 
such as $B\to (\phi,\eta',\pi^0, \omega, K_sK_s,.
..)K_s$ also allow an experimental determination of $\sin 2\beta$ in the
SM~\cite{GW97,LS97}. 
This method is less clean as it has some hadronic uncertainty,
which was naively
estimated to be at the level of $5\%$ \cite{LS97, GWI97}. 
Unfortunately, this 
uncertainty cannot
be reliably determined in a model-independent manner. However, several 
different estimates \cite{LS24-27} find that amongst these modes, 
($\phi$, $\eta'$, $K_s$$K_s$)$K_s$ are rather clean up to an error of only a 
few percent. In passing, we note also another  
intriguing feature of many such penguin-dominated modes is that the central 
value of $\sin 2\beta$ that they give seems to be 
below the two SM predicted values given above in \#1 and in fact,
in many cases, even below 
the value measured via $B\to\psi K_S$ (given in \#2).}

\item{ Another apparent difficulty for the SM is understanding the rather large
difference in the direct {\it CP} asymmetries $\Delta A_{CP} \equiv 
A_{CP}(B^-\to K^-\pi^0) - A_{CP}(\bar{B^0}\to K^-\pi^+)= (14.4\pm 2.9)\%$ 
\cite{HFAG}. Naively this difference is supposed to be zero. Using QCD 
factorization \cite{BBNS} in conjunction with any of the four 
scenarios for $1/m_b$ corrections that have been proposed \cite{BN03} we were 
able to estimate $\Delta A_{CP} = (2.5 \pm 1.5)\%$ \cite{LS07} which is 
several $\sigma$'s away from the experimental observations. It is important to 
understand that by varying over those four scenarios one is actually spanning 
the space of a large class of final state interactions; therefore the 
discrepancy with experiment is serious \cite{FJPZ}. } However, given our limited
understanding of hadronic decays makes it difficult to draw compelling
conclusions from this difficulty for the SM3.

\item{ Finally, more recently the possibility of the need for a largish 
non-standard {\it CP}-phase has been raised \cite{UT08,uli3} in the study of 
$B_s\to \psi\phi$
at Fermilab by CDF \cite{cdf} and D0 \cite{D0} experiments. Since the above items suggest the 
presence of a beyond the SM {\it CP} odd phase in $b\to s$ transitions as 
(for example ) already emphasized in \cite{LS07}, such non-standard effects in
$B_s$ decays are quite unavoidable.}

\end{enumerate}
   
In the following we show that SM with a fourth generation [SM4] 
is readily able to address these difficulties and in particular the data seems 
to suggest the need for $m_{t'}$ within the range (400 - 600) GeV. 

SM4 is a simple extension of SM3 with additional up-type (t')  and
down-type ( b' ) quarks.  It retains all the features of the SM.  The
t'  quark like u, c, t quarks contributes in the $b \to s$ transition at
the loop level  \cite{AS_olds1}. The addition of fourth generation means that the quark
mixing matrix will become a $4 \times 4$ matrix and the parametrization of
this unitary matrix requires six real parameters and three phases. 
The two extra phases imply the possibility
of extra sources of {\it CP} violation \cite{AS_olds3}. In order to find out the 
limits on these extra parameters along with the other observables we 
concentrate mainly on the constraints that will come from $B_d - \bar{B_d}$ and
 $B_s - \bar{B_s}$ mixing, ${\cal{BR}}(B\to X_s \gamma)$, ${\cal{BR}}(B\to X_s 
\ell^+ \ell^-)$ \cite{highq2},
indirect {\it CP} violation in $K_L \to \pi\pi$ described 
by $|\epsilon_k|$ etc. Table I summarizes complete list of inputs 
that we have used to constrain the SM4 parameter space. With these input 
parameters we have made the scan over the entire parameter space by a flat 
random number generator and obtained the constraints on various parameters of 
the 4$\times$4 mixing matrix. In Table-II we present the one sigma allowed 
ranges of $|V^{\ast}_{t's}V_{t'b}|$ and $\phi_s'$ (the phase of
$V_{t's}$), which follow from our
analysis~\cite{MC09}.

\begin{table}[t]
\begin{center}
\begin{tabular}{|l|}
\hline
$B_K = 0.72 \pm 0.05$ \\
$f_{bs}\sqrt{B_{bs}} = 0.281 \pm 0.021$  GeV \\
$\Delta{M_s} = (17.77 \pm 0.12) ps^{-1}$\\
$\Delta{M_d} = (0.507 \pm 0.005) ps^{-1}$ \\
$\xi_s = 1.2 \pm 0.06$\\
$\gamma = (75.0 \pm 22.0)^{\circ} $ \\
$|\epsilon_k|\times 10^{3} = 2.32 \pm 0.007$\\
$\sin 2\beta_{\psi K_s} = 0.672 \pm 0.024$\\
${\cal{BR}}(K^+\to \pi^+\nu\nu) = (0.147^{+0.130}_{-0.089})\times 10^{-9}$\\
${\cal{BR}}(B\to X_c \ell \nu) = (10.61 \pm 0.17)\times 10^{-2}$\\
${\cal{BR}}(B\to X_s \gamma) = (3.55 \pm 0.25)\times 10^{-4}$\\
${\cal{BR}}(B\to X_s \ell^+ \ell^-) = (0.44 \pm 0.12)\times 10^{-6}$ \\
( High $q^2$ region )\\
$R_{bb} = 0.216 \pm 0.001$\\
$|V_{ub}| = (37.2 \pm 5.4)\times 10^{-4}$\\
$|V_{cb}| = (40.8 \pm 0.6)\times 10^{-3} $\\
$\eta_c = 1.51\pm 0.24$ \cite{uli1} \\
$\eta_t = 0.5765\pm 0.0065$ \cite{buras1}\\
$\eta_{ct} = 0.47 \pm 0.04$ \cite{uli2}\\
$m_t = 172.5$ GeV\\
\hline
\end{tabular}
\caption{Inputs used to constrain the SM4 parameter space; 
the error on $V_{ub}$ is increased to reflect the
disagreement between the inclusive and exclusive methods.}
\label{tab1}
\end{center}
\end{table}

The SM3 expressions for $\epsilon_k$ and $Z\to b\bar{b}$ decay width have been
taken from \cite{buras2} and \cite{zbb} respectively whereas the relevant
expressions for $\Delta{M_s}\over \Delta{M_d}$, along with the other
observables can be found in \cite{buras3}. The corresponding expressions in 
SM4 i.e., the additional contributions arising due to $t'$ quark can be 
obtained by replacing the mass of $t$-quark by $m_{t'}$ in the respective 
Inami-Lim functions. 
For concreteness, we use the parametrization
suggested 
%by Botella and Chau 
in \cite{chau} for 4$\times$4 CKM matrix 
[$V_{CKM4}$]. In $\Delta{M_d}$ and $\Delta{M_s}$, apart from the other factors, we have the CKM 
elements $V_{tq}V^{\ast}_{tb}$ which can be replaced by 
%and  $\Delta{M_s}$ respecitively, 
%ontains the term
%$V_{ts}V^{\ast}_{tb}$ 
(with q=d or s),
\bea
V_{tq}V^{\ast}_{tb} &=& -(V_{uq}V^{\ast}_{ub} + V_{cq}V^{\ast}_{cb}
+V_{t'q}V^{\ast}_{t'b} ) 
\nonumber \\
%V_{ts}V^{\ast}_{tb} &=& -(V_{us}V^{\ast}_{ub} + V_{cs}V^{\ast}_{cb}
%+V_{t's}V^{\ast}_{t'b} ),
\label{unitary}
\eea
%which is replaced by 
using the 4$\times$4 CKM matrix unitarity relation, 
$\lambda_u+\lambda_c+\lambda_t+ \lambda_{t'}=0$
where $\lambda_q= V_{qb}V_{qs}^*$.    
%\noindent using the 4$\times$4 CKM matrix unitarity relations. 
The phase of 
$V_{td}$ and $V_{ts}$ will also be obtained by using this unitarity
relation.  
In this way we can reduce the number of unknown parameters by 
using information from known parameters. 
%We have also used the unitarity 
%relations for $|V_{tb}|^2$ in the expression for $\Gamma(Z\to b\bar{b})$.

With a sequential fourth generation, the effective Hamiltonian
describing the $b \to s$ transitions
% $B^- \to \pi^0 K^-
%$ and $\bar B^0 \to \pi^+ K^-$
becomes 
\bea 
{\cal H}_{eff} &=&\frac{G_F}{\sqrt{2}}\big[ \lambda_u(C_1^u O_1+C_2^u 
O_2  +\sum_{i=3}^{10} C_i^u O_i) \nonumber \\
&{}& + \lambda_c\sum_{i=3}^{10} C_i^c O_i
-\lambda_{t'}\sum_{i=3}^{10} \Delta C_i^{t'} O_i\big] \;, 
\eea
%where $\lambda_q= V_{qb}V_{qs}^*$ and we have used  the unitarity
%condition $\lambda_u+\lambda_c+\lambda_t+ \lambda_{t'}=0$. 
where $C_i^q$'s
are the Wilson coefficients, $\Delta C_i^{t'}$'s  are the effective
(t subtracted) $t'$ contributions and $O_i$ are the current-current
operators. 
Using the above Hamiltonian, 
%with a sequential fourth generation 
%\cite{CHOU}, that describes the $b\to s$ transition 
and
following \cite{LS07} we use the S4 scenario of QCD factorization 
approach  \cite{BN03}
for the evaluation of hadronic matrix elements  and the amplitudes
for the decay modes $B\to \pi K$ and $B\to \phi K_s$ for  
$m_{t'}= 400$ , $500$ and $600$ GeV respectively. 

Using the ranges of $\lambda^{s}_{t'} \equiv |V^{\ast}_{t's}V_{t'b}|$  and 
$\phi_s'$, 
%the phase of $V_{t's}$ 
as obtained from the fit for different 
$m_{t'}$ (Table II), we studied the allowed regions in the 
$\Delta A_{CP}-\lambda^{s}_{t'}$ plane for different values of $m_{t'}$.
%as 
%shown in Fig.{\ref{fig1}}.  
With the 4th family we see that there is some enhancement and $\Delta A_{CP}$ up to about 8\% may be feasible which is till
somewhat small compared to the observed value ($14.4 \pm 2.9$)\%. Again,
as we mentioned this could be due to the inadequacy of the QCD factorization model we are using.

\begin{table}[htbp]
\begin{center}
\begin{tabular}{|l|c|c|l|}
\hline
$m_{t'}$&400 & 500 & 600 \\
\hline
$\lambda^s_{t'}$ & (0.08 - 1.4)& (0.06 - 0.9)& (0.05 - 0.7) \\
\hline
$\phi_s'$ & -80 $\to$ 80 & - 80 $\to$ 80 & -80 $\to$ 80 \\
\hline
\end{tabular}
\caption{Allowed ranges for the parameters, $\lambda^s_{t'}$ ($\times 10^{-2}$)
and phase $\phi_s'$ (in degree) for different masses $m_{t'}$ ( GeV), that has
been obtained from the fitting with the inputs in Table I.}
\label{tab2}
\end{center}
\end{table}

%\begin{figure}
%\includegraphics[width= 0.80 \linewidth]{acp.eps}
%\hspace{3.3cm}
%\caption{The allowed range of the CP asymmetry difference ($\Delta A_{CP}$) in
%the ($\Delta A_{CP}-\lambda^{s}_{t'}$) plane, where the red, green, magenta
%and blue regions correspond to $m_{t'}$ = 400, 500, 600 and 700 GeV. The 30
%\% error bars due to hadronic uncertainties \cite{LS07} are shown by grey
%bands. The balck and red horizontal lines correspond to the experimentally 
%allowed 1 and 2-$\sigma$ range respectively.}
%\label{fig1}
%\end{figure}

\begin{figure}[htbp]
\parbox{8cm}{
\epsfig{file=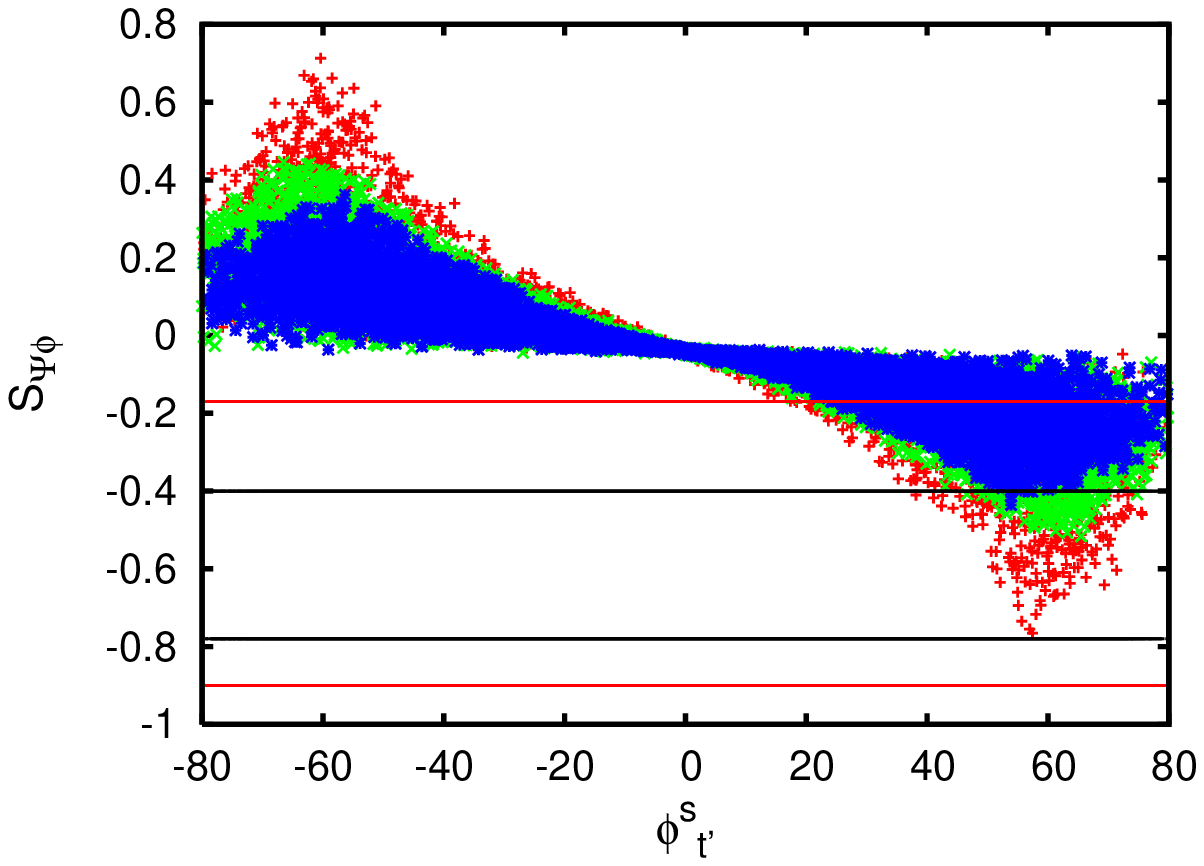,width=8cm,angle=0}
}
\parbox{8cm}{
\epsfig{file=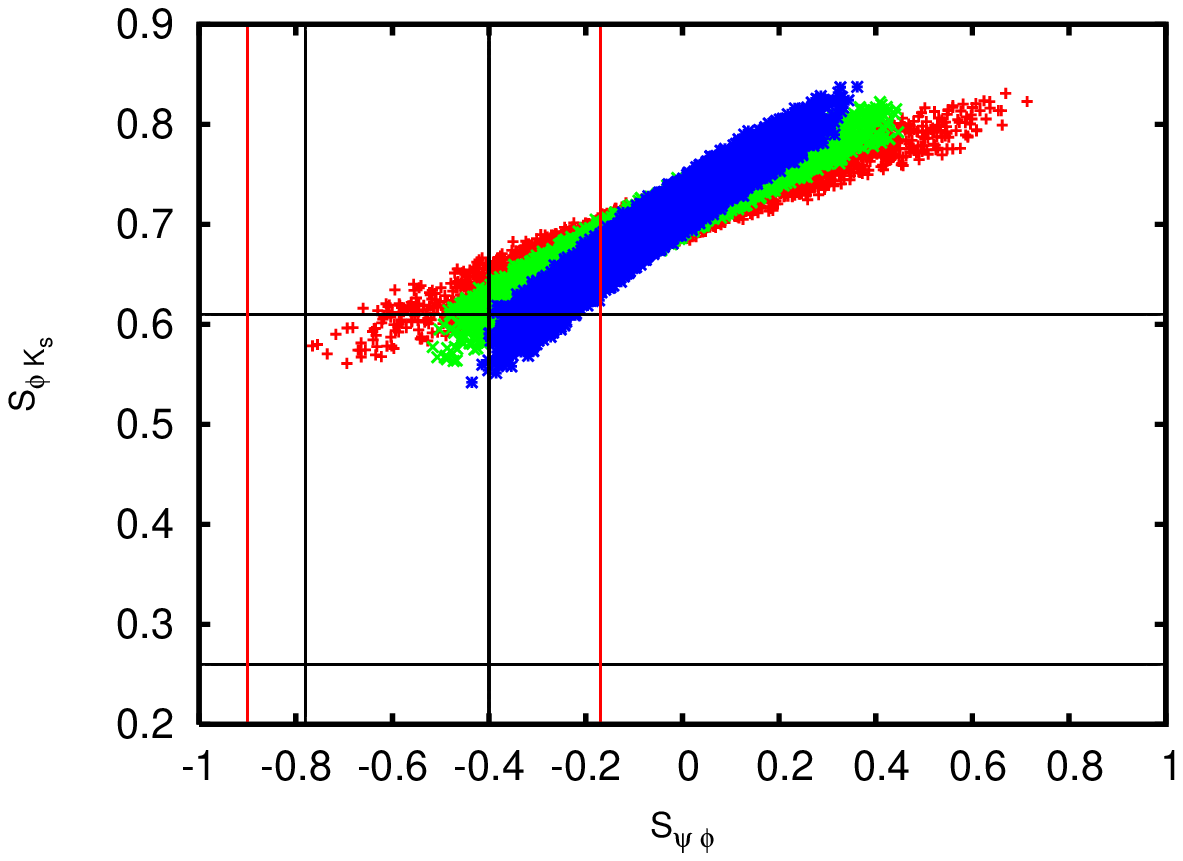,width=8cm,angle=0}
}
\caption{
The left panel shows the allowed range for $S_{\psi \phi}$ in the 
($S_{\psi\phi}-\phi^{s}_{t'}$) plane for $m_{t'}= 400$ (red), $500$ (green) 
and $600$ (blue)
%and $700$ (blue) 
GeV respectively. 
Black and red horizontal 
lines  in the figure indicate 1-$\sigma$ and 2-$\sigma$ experimental ranges 
for $S_{\psi\phi}$ respectively.
The right panel shows the correlation between $S_{\phi K_s}$ and $S_{\psi\phi}$
for $m_{t'}= 400$ (red), $500$ (green) and $600$ (blue) 
%and $700$ (blue) 
GeV
respectively. The horizontal lines represent the experimental
$1\sigma$ range for $S_{\phi K_s}$ whereas the vertical lines (Black
1-$\sigma$ and red 2-$\sigma$ ) represent that for $S_{\psi\phi}$ .
\label{fig1}}
\end{figure}

%\begin{figure}[t]
%\includegraphics[width= 0.80 \linewidth]{s_bs.eps}
%\caption{The allowed range for $S_{\psi \phi}$ in the ($S_{\psi\phi}-\phi^{s}_
%{t'}$) plane for $m_{t'}= 400$ (red), $500$ (green), $600$ (magenta) and $700$
% (blue) GeV respectively. Black and red horizontal lines  in the figure 
%indicate 1-$\sigma$ and 2-$\sigma$ experimental ranges for $S_{\psi\phi}$
%respectively.}
%\label{fig2}
%\end{figure}

%\begin{figure}
%\includegraphics[width= 0.80 \linewidth]{scp.eps}
%\caption{Correlation between $S_{\phi K_s}$ and $S_{\psi\phi}$
%for $m_{t'}= 400$ (red), $500$ (green), $600$ (magenta) and $700$ (blue) GeV 
%respectively. The horizontal lines represent the experimental  
%$1\sigma$ range for $S_{\phi K_s}$ whereas the vertical lines (Black 
%1-$\sigma$ and red 2-$\sigma$ ) represent that for $S_{\psi\phi}$ . }
%\label{fig3}
%\end{figure}

In Fig.\ref{fig1} (left-panel) we have shown the allowed regions in the $S_{\psi\phi}-
\phi^{s}_{t'}$ plane for different values of $m_{t'}$ and
in the right-panel of Fig.\ref{fig1} we have shown the correlation between
{\it CP} asymmetries in $B\to \psi\phi$ and $B\to \phi K_s$. We follow the 
notation $S_{\psi \phi} = \sin(\phi^{\Delta}_s -2\beta_s) = \sin2\beta^{eff}_s$
, where $\phi^{\Delta}_s$ is the phase coming from mixing and
$\beta_s = arg(- {V^{\ast}_{tb}V_{ts}\over V^{\ast}_{cb}V_{cs}}) = 1.1^{\circ}
\pm 0.3^{\circ}$, is the phase of $b\to c \bar{c} s$ decay amplitude
\cite{beneke1, uli3}. The range for new $B_s$ mixing phase $\phi^{\Delta}_s$ 
is given (@68\% CL) by 
%\bea%%
$\phi_s^\Delta \,\in \, \lt( -18 \pm 7 \rt)^{\circ}$   
%\qquad\mbox{or }\qquad$
or,     
%\nonumber \\
$\phi_s^\Delta \, \in \,  \lt( -70 \pm 7 \rt)^{\circ}$.    
% \qquad
% @68\% \, \mbox{CL} .
%\label{cdfphi}
%\eea%%
The corresponding 2-$\sigma$ and 1-$\sigma$ ranges for $S_{\psi\phi}$ is given
by [-0.90, -0.17] and [-0.78, -0.40] respectively. The large
error on $S_{\phi K_s}$ and $S_{\psi\phi}$ does not allow at present to draw
strong conclusions on $m_{t'}$, nevertheless the present experimental bounds
disfavor $m_{t'}>$  600 GeV.

A very appealing feature of the 4th family hypothesis is that it
rather naturally
explains the pattern of the observed anomalies.  First of all
the heavy $m_{t'}$ generates a very important new source of electroweak penguin 
(EWP) contribution since, as is well known, these amplitudes are able to avoid 
the decoupling theorem and grow as $m_{t'}^2$ \cite{IL,AS_olds1}. This helps to 
explain two of the anomalies in $b \to s$ transitions. The enhanced EWP 
contribution helps in explaining the difference in CP-asymmetries, $\Delta A_{CP}$ as it is really the $K^{\pm} \pi^0$ that is enhanced because of the color 
allowed coupling of the Z to the $\pi^0$.
%\cite{SM4pi0}. 
A second important 
consequence of $t'$ is that  $b \to s$ penguin has a new CP-odd phase carried 
by $V_{t'b}V^*_{t's}$. This is responsible for the fact that $\sin 2 \beta$ 
measured in $B \to \psi K_s$ differs with that measured in penguin-dominated 
modes such as $B \to (\phi, \eta', K_s K_s…)K_s$.

Note also that $\Delta B=2$ box graph gets important new contributions
from the $t'$
since these amplitudes as mentioned before are proportional to
$m_{t'}^2$.  Furthermore,
they are accompanied by new CP-odd phase which is not present in SM3. 
This phase is responsible for the fact that the $\sin 2\beta$ measured in 
$B \to\psi K_s$ is lower than the value(s) ``predicted" in SM3 \cite{LS08}
given in item \# 1 on page 1.

Finally, we note briefly in passing how SM4 gives a very simple
explanation for the size of the new CP-phase effects in $B_d$ versus $B_s$ 
mesons. In $B_d$ oscillations resulting in $ B \to \psi K_s$, top quark plays 
the dominant role and we see that the measured value of $\sin2 \beta$ deviates 
by $\approx 15\%$ from predictions of SM3.  It is then the usual hierarchical 
structure of the mixing matrix (now in SM4) that guarantees that on 
$\sin2 \beta$, $t'$ will only have a subdominant effect.
However,  when we consider $B_s$ oscillations then the role
of $t'$ and $t$ get reversed. In $B_s$ mixing the top quark in SM3 has
negligible CP-odd phase. Therein then the $t'$ has a pronounced effect. SM4 
readily explains that just as $t$ is dominant in $\sin 2 \beta$ and 
subdominant in $\sin2 \beta_s$, the $t'$ is dominant in $\sin 2 \beta_s$ and  
subdominant in $\sin 2 \beta$.

We now briefly summarize some of the definitive signatures of the 4th
family scenario in flavor observables~\cite{WIP}. The need for new CP phase(s) 
beyond the single KM phase\cite{1973} of course must continue to persist. 
This means
that the three values of $\sin 2 \beta$, the fitted one, the one
measured
via $\psi K_s$ and the one measured via penguin dominated modes
({\it e.g.} $\phi K_s$, $\eta'K_s$ etc.) should continue to 
differ from each other as more accurate analyses become available.  
Furthermore, $B_s$ mixing should also continue to 
show the presence of a non-standard phase ({\it e.g.} in $B_s -> \psi
\phi$) as higher statistics are accumulated.          
For sure SM4 will have many more interesting
applications in flavor physics which need to be explored.  For the LHC,
one definitive pediction of this analysis is a $t'$ with $m_{t'}$
in the range of $\approx$ 400-600 GeV and the   
detection of the $t'$, $b'$ and their leptonic counterparts deserves attention.
EW precision constrains the mass-splitting between $t'$ and $b'$ to be small,
around $50$ GeV \cite{NEED,NORV02}. 
%This constraint will clearly have important 
%repercussions for their decays and therefore their signals at the LHC. 

As far 
as the lepton sector is concerned, it is clear that the 4th family leptons have
 to be quite different from the previous three families in that the neutral 
leptons have to be rather massive, with masses $> m_Z/2$. This may also be a 
clue that the underlying nature of the 4th family may be quite different
from the previous three families; for one thing it could be relevant to the dark matter issue\cite{DM4}. It may also open up the possibility of unification with
the SM gauge group~\cite{PQH98}. KM \cite{1973} mechanism taught us 
the crucial role 
of the three families in endowing CP violation in SM3. It is conceivable that 
4th family plays an important role \cite{HOU08,CJ88,FDA96,GK08} in yielding enough CP to 
generate baryogenesis which is difficult in SM3. Of course it also seems highly
 plausible that the heavy masses in the 4th family play a significant role in 
dynamical generation of electroweak symmetry breaking. In particular, the 
masses around 500 or 600 GeV that are being invoked in our study, point to a 
tantalizing possibility of dynamical electroweak symmetry breaking as the 
Pagels-Stokar relation in fact requires quarks of masses around 500 or 600 GeV 
for dynamical mass generation to take place \cite{PS,HHT,BH,BURD}.
Note also that for such heavy masses the values of Yukawa coupling will be large %but not necessarily so
%large to break down 
so that corrections to perturbation theory may not be negligible~\cite{uni}.
Finally, we want to emphasize that a fourth family of quarks 
does {\it not} violate electro-weak precision tests~\cite{KPST07}.  
Clearly all this brief discussion is signaling is that there is a lot of 
physics involving the new family that needs to be explored and understood. 

We thank Graham Kribs, Ulrich Nierste and Marc Sher for discussions.
The work of AS is supported in part by the US DOE contract No. 
DE-AC02-98CH10886. This research grew as a part of a working group on flavor
physics at the $X^{th}$ Workshop of High Energy Physics 
Phenomenology (WHEPP X), Chennai,
India (Jan 08). AS is grateful to organizers especially to Rahul Basu and
Anirban Kundu for the invitation. The work of RM is supported by DST and AG by 
CSIR, Government of India. SN's work is supported in part by
MIUR under contact 2004021808-009 and by a European Community's
%Marie-Curie
%Research Training Network under 
contract MRTN-CT-2006-035505.
%``Tools and
%Precision Calculations for Physics Discoveries at Colliders".


\begin{thebibliography} {99}

\bibitem{1963}
N. Cabibbo, \PRL(10,531,1963).
%%CITATION = PRLTA, 10,531;%%
\bibitem{1973}
M. Kobayashi and T. Maskawa, Prog. Theor. Phys. {\bf 49}, 652 (1973).
%%CITATION = PTPKA,49,652;%%
\bibitem{HFAG}
See http://www.slac.stanford.edu/xorg/hfag/ 

\bibitem{RMP}
T. E. Browder {\it et al}, arxiv:0802.3201 .
%%CITATION = ARXIV:0802.3201;%%

\bibitem{LS07}
E. Lunghi and A. Soni, \JHEP(0709,053,2007).   
%%CITATION = JHEPA, 0709,053;%%

\bibitem{LS08}
E. Lunghi and A. Soni, arxiv:0803.4340 .
%%CITATION = ARXIV:0803.4340;%%


%\bibitem{UT07}
%M. Bona {\it et al.} [UT Fit collaboration], arxiv:0707.0636 .
%%CITATION = JHEPA, 0803,049;%%


\bibitem{AS_olds1}
The importance of rare B-decays for searching for the 4th family was
emphasized long ago in
W. -S. Hou, 
R. Willey and A.Soni, Phys. Rev. Lett. {\bf 58}, 1608 (1987);
see also~\cite{AS_olds2,AS_olds3,AS_olds4}.
%%CITATION = PRLTA, 58,1608;%%
\bibitem{AS_olds2}
W. -S. Hou,
A. Soni and H. Steger, Phys. Rev. Lett. {\bf 59}, 1521 (1987).
%%CITATION = PRLTA, 59,1521;%%
\bibitem{AS_olds3}
W. -S. Hou,
A. Soni and H. Steger, \PL(B192,441,1987).
%%CITATION = PHLTA,B192,441;%% 
\bibitem{AS_olds4}
C. Hamzaoui, A. I. Sanda and A. Soni,  
Phys. Rev. Lett. {\bf 63}, 128 (1989).

\bibitem{hung-sher}
For a general review on the 4th family see,
P. H. Frampton, P. Q. Hung and M. Sher,
Phys. Rept. {\bf330},363,(2000).


\bibitem{CHOU}
Possible role of the fourth family in $B$-decays has also been emphasized in,
W. -S. Hou {\it et al.}, \JHEP(0609,012,2006);
%M. Nagashima, G. Raz and A. Soddu, \JHEP(0609,012,2006);
%%CITATION = JHEPA, 0609,012;%%
W. -S. Hou {\it et al.}, \PRL(95,141601,2005);
%M. Nagashima and A. Soddu, \PRL(95,141601,2005);
%%CITATION = PRLTA, 95,141601;%%
W. -S. Hou {\it et al.}, \PRL(98,131801,2007);
%[hep-ph/061107]; 
%H. -Nan Li, S. Mishima and M. Nagashima, \PRL(98,131801,2007) [arxiv: hep-ph/061107].
%%CITATION = PRLTA, 98,131801;%%
%W. -S. Hou {\it et al.}, 
\PRD(76,016004,2007).
%[hep-ph/0610385].
We extend and complement these works and provide specific constraints on the 
relevant parameters especilly on $m_{t'}$.

\bibitem{error_etc}
We want to briefly mention that in an earlier version of this
manuscript we had made an error in our analysis and also we are now using
somewhat different input to constrain the four generation SM compared to what we had done before. The result is that the data no longer seems to require masses
greater than 600 GeV.
\bibitem{GW97}
Y. Grossman and M. Worah, \PL(B395,241,1997) [hep-ph/9612269].
%%CITATION = PHLTA,B395,241;%%
\bibitem{LS97}
D. London and A. Soni, \PL(B407,61,1997) [hep-ph/9704277].
%CITATION = PHLTA,B407,61;%%

\bibitem{GWI97}
Y. Grossman,  
G. Isidori and M. Worah, \PRD(58,057504,1998) [arxiv:hep-ph/9708305].
%%CITATION = PHRVA,D58,057504;%%

\bibitem{LS24-27} {\it et al.}, 
M. Beneke, \PL(B620,143,2005);
%[hep-ph/0505075]; 
%%CITATION = PHLTA,B620,143;%%
H. -Y. Cheng,
C. -K. Chua and A. Soni, \PRD(72,014006,2005).
%%CITATION = PHRVA,D72,014006;%%
\PRD(72,094003,2005); %%CITATION = PHRVA,D72,094003;%%
G. Buchalla,
G. Hiller, Y. Nir and G. Raz, \JHEP(0509,074,2005);
%%CITATION = JHEPA, 0509,074;%%
A. Williamson and J. Zupan, \PRD(74,014003,2006).
%[hep-ph/0601214].
%%CITATION = PHRVA,D74,014003;%%
\bibitem{BBNS}M. Beneke, 
G. Buchalla, M. Neubert and C. Sachrajda, \NP(B606,245,2001); [hep-ph/0104110].
\bibitem{BN03} M. Beneke and M. Neubert, \NP(B675,333,2003).
%[hep-ph/0308039].
%%CITATION = NUPHA,B675,333;%%
\bibitem{FJPZ}
See also, R. Fleischer,
% {\it et al.},  arxiv:0806.2900. 
S. Jager, D. Pirjol and J. Zupan, arxiv:0806.2900 [hep-ph].
%%CITATION = ARXIV:0806.2900;%%

\bibitem{uli3}
A.~Lenz and U.~Nierste, \JHEP(0706,072,2007).    
%  [hep-ph/0612167].
  %%CITATION = HEP-PH/0612167;%%

\bibitem{UT08}
M. Bona {\it et al.} [UTFit collaboration],
arxiv:0803.0659.
%%CITATION = ARXIV:0803.0659;%%
\bibitem{cdf}
T. Aaltonen {\it et al.} [CDF Collaboration], arxiv:0712.2397.
%%CITATION = ARXIV:0712.2397;%%
\bibitem{D0}
V.~M.~Abazov {\it et al.} [D0 Collaboration], arxiv:0802.2255.
%%CITATION = ARXIV:0802.2255;%%

\bibitem{uli1}
S. Herrlich and U. Nierste, \NP(B419,292,1994).
%[hep-ph/9310311].
%%CITATION = NUPHA,B419,292;%%

\bibitem{buras1}
A. J. Buras, 
M. Jamin and P. H. Weisz, \NP(B347,491,1990).
%%CITATION = NUPHA,B347,491;%%
\bibitem{uli2}
S. Herrlich and U. Nierste, \PRD(52,6505,1995).
%[hep-ph/9507262].
%%CITATION = PHRVA,D52,6505;%%

\bibitem{MC09}
Recently coonstraints due to oblique corrections have also been included see,
M. S. Chanowitz, arXiv:0904.3570 [hep-ph]; see also~\cite{WIP}. As noted by Chanowitz, although these constraints eliminate a portion of the
parameter space, a large fraction still remains valid. 

\bibitem{WIP} More details will be given in a forthcoming article.



\bibitem{highq2}
%The measurements of the $B\to X_s\ell^+ \ell^-$ in the
%two regions, so called low $q^2$ $(q^2\protect\lsim 6 GeV^2)$  and high $q^2$ 
%$(q^2\protect\gsim 14 {GeV}^2)$, are complementary as they have different 
%sensitivities to the short distance physics. 
Since in the low $q^2$ region the 
photon contribution to $B_s\to X_s\ell^+\ell^-$ is likely to be dominant and
since it is the $Z$ that is very sensitive to $m_{t'}$, as the
Z-exchange  amplitude grows as $m^2_{t'}$ (see 
%the first paper of 
~\cite{AS_olds1}), we are only using the Br in the high $q^2$ region to constrain SM4.
However, we checked that use of low $q^2$ region does not change things
much.
\bibitem{buras2} A. J. Buras, D. Guadagnoli, arxiv:0805.3887.
%%CITATION = ARXIV:0805.3887;%%
\bibitem{zbb}
J. Bernabeu,
% {\it et al.}, \NP(B363,326,1991).
A. Pich and A. Santamaria, \NP(B363,326,1991).
%%CITATION = NUPHA,B363,326;%%
\bibitem{buras3}
G. Buchalla and A. J. Buras, \NP(B548,309,1999);
%[hep-ph/9901288];
%%CITATION = NUPHA,B548,309;%%
A. J. Buras and R. Fleischer, Adv.\ Ser. \ Direct.\ High Energy Phys.\ {\bf 15},
65 (1998);
%[hep-ph/9704376]; 
%%CITATION = 00319,15,65;%%
A. J. Buras, hep-ph/9806471.
%%CITATION = HEP-PH/9806471;%%
\bibitem{chau}
F. J. Botella and L. L. Chau, \PL(B168,97,1986).
%%CITATION = PHLTA,B168,97;%%
%\bibitem{BN03} M. Beneke and M. Neubert, \NP(B675, 333, 2003) [arxiv: %hep-ph/0308039].
%%CITATION = NUPHA,B675,333;%%
\bibitem{beneke1}
M.~Beneke {\it et al.},
%G.~Buchalla, A.~Lenz and U.~Nierste,
%``CP asymmetry in flavor specific B decays beyond leading logarithms,''
Phys.\ Lett.\ B {\bf 576} (2003) 173.   
%[hep-ph/0307344].
%%CITATION = PHLTA,B576,173;%%
M.~Ciuchini {\it et al.},
%E.~Franco, V.~Lubicz, F.~Mescia and C.~Tarantino,
%``Lifetime differences and CP violation parameters of neutral B mesons at the
%next-to-leading order in QCD,''
\JHEP(0308,031,2003).       
%[hep-ph/0308029].
%%CITATION = JHEPA,0308,031;%%


%\bibitem{uli3}
%A.~Lenz and U.~Nierste, \JHEP(0706,072,2007).    
%  [hep-ph/0612167].
  %%CITATION = HEP-PH/0612167;%%

\bibitem{IL} T. Inami and C. S. Lim, Prog. Theor. Phys. {\bf 65},
297 (1981); {\it ibid} {\bf 65}, 1772E (1981);
%%CITATION = PTPKA,65,297;%%
see also, G. Buchalla {\it et al.}, Rev. Mod. Phys. 68: 1125, 1996.
%[hep-ph/9512380].
%%CITATION = RMPHA,68,1125;%% 
%\bibitem{SM4pi0}
%Clearly due to enhanced EWP coupling to $\pi^0$ in SM4, $B\to \pi^0\pi^0$ 
%should also receive appreciable contribution.


\bibitem{NEED}
Particle Data Group ( D. E. Groom {\it et al}), \EPJC(15,1,2000);
M. Maltoni {\it et al.}, \PL(B476,107,2000).
%[hep-ph/9911535];
%%CITATION = PHLTA,B476,107;%%\PL(B529,111,2002
\bibitem{NORV02}
V. A. Novikov, 
%[hep-ph/0111028].
L. B. Okun, A. N. Rozanov and M. I. Vysotsky, \PL(B529,111,2002)[arxiv: hep-ph/0111028].
%%CITATION = PHLTA,B529,111;%%

\bibitem{DM4}
See, {\it e.g.}
%\cite{Volovik:2003kh}
  G.~E.~Volovik,
    %``Dark matter from SU(4) model,''
      Pisma Zh.\ Eksp.\ Teor.\ Fiz.\  {\bf 78}, 1203 (2003)
        [JETP Lett.\  {\bf 78}, 691 (2003)].  
%	  [hep-ph/0310006].
	    %%CITATION = JTPLA,78,691;%%


\bibitem{PQH98}
P. Q. Hung, \PRL(80,3000,1998).

\bibitem{HOU08}
W. S. Hou, arxiv:0803.1234. For earlier related works see, \cite{CJ88,FDA96} 
%%CITATION = ARXIV:0803.1234;%%

\bibitem{CJ88}
C. Jarlskog and R. Stora, \PL(B208,288,1988).

\bibitem{FDA96}
F. del Aguila and J. A. Aguilar-Saavedra, \PL(B386,241,1996);
F. del Aguila and J. A. Aguilar-Saavedra and G. C. Branco,Nucl. Phys. {\bf B510},39,1998.


\bibitem{GK08}
See also,
R.~Fok and G.~D.~Kribs,
  %``Four Generations, the Electroweak Phase Transition, and
    %Supersymmetry,''
        arXiv:0803.4207 .
	      %%CITATION = ARXIV:0803.4207;%%

\bibitem{PS}
H. Pagels and S. Stokar, \PRD(20,2947,1979).
%%CITATION = PHRVA,D20,2947;%%

\bibitem{HHT}
H. -J. He, 
%[hep-ph/0108041].
C. T. Hill and T. M. P. Tait, \PRD(65,055006,2002) [arxiv:hep-ph/0108041].
%%CITATION = PHRVA,D65,055006;%%
\bibitem{BH}
B. Holdom, \JHEP(0608,076,2006).
%[hep-ph/0606146].
%%CITATION = JHEPA,0608,076;%%
\bibitem{BURD}
G. Burdman and L. D. Rold, \JHEP(0712,086,2007).
%arxiv:0710.0623 .
%%CITATION = JHEPA,0712,086;%%
\bibitem{uni}
For unitarity issues 
%pertaining to such heavy quarks, 
see:
M. S. Chanowitz, \PL(B352,376,1995) 
%[hep-ph/9503458] 
and
M. S. Chanowitz {\it et al.},
\PL(B78,285,1982).
%\bibitem{yukawa}
%Such heavy fermion masses do entail large Yukawa couplings, therefore
%higher order corrections to the estimates presented here 
%may be sizeable. 
 
\bibitem{KPST07}
%For electroweak precision tests  
%see: 
G. D. Kribs, T. Plehn, M. Spannowsky and T. M. P. Tait,   
\PRD(76,075016,2007). 
%arxiv: 0706.3718. In fact
%This study shows that with $m_{t'} \gsim$ 400 GeV,
%the Higgs has to be heavier than about 300 GeV.
%%CITATION = PHLTA,B78,285;%%
%%CITATION = PHLTA,B352,376;%%
%%CITATION = PHRVA,D76,075016;%%


\end{thebibliography}
\end{document}